\documentclass[10pt,journal,cspaper,compsoc]{IEEEtran}
\usepackage{times,amsmath,epsfig}
\usepackage{listings}
\usepackage{latexsym}
\usepackage{graphicx}
\usepackage{float}
\usepackage{custom}
\usepackage{enumerate}
\usepackage{color}
\usepackage{mdwlist}

\ifCLASSOPTIONcompsoc
\else
\fi
\ifCLASSINFOpdf
\else
\fi
\begin{document}
%
\title{Program Transformations for Asynchronous and Batched Query Submission}
%
%
%
%

\author{Karthik~Ramachandra, Mahendra~Chavan, Ravindra~Guravannavar, S~Sudarshan


\IEEEcompsocitemizethanks{
\IEEEcompsocthanksitem Karthik Ramachandra, IIT Bombay.\protect\\
E-mail: karthiksr@cse.iitb.ac.in

\IEEEcompsocthanksitem Mahendra Chavan, IIT Bombay (current affiliation: SAP, Pune)\protect\\
E-mail: mahendra.chavan@sap.com

\IEEEcompsocthanksitem Ravindra Guravannavar, IIT Hyderabad (current: independent consultant)\protect\\
E-mail: ravig@acm.org
\IEEEcompsocthanksitem S Sudarshan, IIT Bombay \protect\\
E-mail: sudarsha@cse.iitb.ac.in
}
\thanks{}}

\IEEEcompsoctitleabstractindextext{%
\begin{abstract}
The performance of database/Web-service backed 
applications can be significantly improved by asynchronous submission of 
queries/requests well ahead of the point where the results are needed,
so that results are likely to have been fetched already when they 
are actually needed.  However, manually writing applications to 
exploit asynchronous query submission is tedious and error-prone.  

In this paper we address the issue of automatically transforming a program 
written assuming synchronous query submission, to one that exploits 
asynchronous query submission.  Our program transformation method is based on 
data flow analysis and is framed as a set of transformation rules.  Our rules 
can handle query executions within loops, unlike some of the earlier work in 
this area.  
We also present a novel approach that, at runtime, can combine multiple
asynchronous requests into batches, thereby achieving the benefits of batching
in addition to that of asynchronous submission.
We have built a tool that implements our transformation techniques 
on Java programs that use JDBC calls; our tool can be extended to handle Web 
service calls.  We have carried out a detailed experimental study on several 
real-life applications, which shows the effectiveness of the proposed 
rewrite techniques, both in terms of their applicability and the
performance gains achieved.
\end{abstract}

%

}

\maketitle

\IEEEdisplaynotcompsoctitleabstractindextext

%
\IEEEpeerreviewmaketitle

\section{Introduction} \label{sec:intro}
In many applications calls made to execute database queries or to invoke Web
services are often the main causes of latency. Asynchronous or non-blocking
calls allow applications to reduce such latency by overlapping CPU operations
with network or disk IO requests, and by overlapping local and remote
computation. Consider the program fragment shown in Example \ref{ex:prog1}.
In the example, it is easy to see that by making a non-blocking call to
the database we can overlap the execution of method {\em foo()} with the
execution of the query, and thereby reduce latency.

Many applications are however not designed to exploit the full potential of
non-blocking calls. Manual rewrite of such applications although possible, is
time consuming and error prone. Further, opportunities for asynchronous
query submission are often not very explicit in the code. For instance, 
consider the program fragment shown in Example \ref{ex:prog2}. In the program,
the result of the query, assigned to the variable {\em partCount}, is 
needed by the statement that immediately follows the statement executing 
the query. For the code in the given form there would be no gain in replacing 
the blocking query execution call by a non-blocking call, as the execution
will have to block on a {\em fetchResult} call immediately after making
the {\em submitQuery} call. It is however possible to transform the given
loop, as shown in \refex{ex:prog2-trans}, and thereby enable asynchronous query 
submission.
\begin{floatexample}
\begin{sqlindent}
\\
r = executeQuery(query1); \\
s = foo(); \>\>  // Some computation not dependent on r \\
bar(r, s);  \>\> // Computation dependent on r and s \\
\\
{\bf Code with Asynchronous Query Submission} \\
handle = submitQuery(query1);  // Non-blocking query submit\\
s = foo();  \\
r = fetchResult(handle); // Blocking call to fetch query result \\
bar(r, s);
\end{sqlindent}
\caption{A simple opportunity for asynchronous query submission}
\label{ex:prog1}
\end{floatexample}

\begin{floatexample}
\begin{sqlindent}
qt = dbCon.prepare(\>\>\>\>``{\bf select} count(partkey) \`(s0) \\
            \>\>\>\>\ \ {\bf from} part {\bf where} p$\_$category=?''); \\
while(!categoryList.isEmpty()) $\{$  \`(s1)\\
\>    category = categoryList.removeFirst();  \`(s2)\\
\>    qt.bind(1, category);  \`(s3)\\
\>    partCount = executeQuery(qt);  \`(s4)\\
\>    sum += partCount;  \`(s5)\\
$\}$ 
\end{sqlindent}
\caption{Hidden opportunity for asynchronous query submission}
\label{ex:prog2}
\end{floatexample}

\eat{
\begin{floatexample}
\begin{sqlindent}
template1 = select count(partkey) from part  \\
    \>\>\>\>where p$\_$category={\bf @category} \\

while(!categoryList.isEmpty()) $\{$ \\
\>    category = categoryList.removeFirst(); \\
\>    q = template1.bind(category); \\
\>    partCount = executeQuery(q); \\
\>    sum += partCount; \\
$\}$ 
\end{sqlindent}
\caption{Hidden opportunity for asynchronous query submission}
\label{ex:q2}
\end{floatexample}
}

\begin{floatexample}
\begin{sqlindent}
qt = dbCon.prepare(\>\>\>\>``{\bf select} count(partkey) \\
            \>\>\>\>\ \ {\bf from} part {\bf where} p$\_$category=?''); \\

int handle[{\small MAX$\_$SIZE}], n=0; \\
while(!categoryList.isEmpty()) $\{$ \\
\>    category = categoryList.removeFirst(); \\
\>    qt.bind(1, category); \\
\>    handle[n++] = submitQuery(qt); \\
$\}$  \\

for(int i = 0; i $<$ n; i++) $\{$ \\
\>    partCount = fetchResult(handle[i]); \\
\>    sum += partCount; \\
$\}$ 
\end{sqlindent}
\caption{Loop Transformation to Enable Asynchronous Query Submission}
\label{ex:prog2-trans}
\end{floatexample}

The rewritten program in \refex{ex:prog2-trans} contains two loops; the first 
loop submits queries in a non-blocking mode and the second loop uses a 
blocking call to fetch the results and then executes the statements that 
depend on the query results. 

The original program is likely to be slow since it makes multiple synchronous 
requests to the database, each of which incurs network round trip delays, as 
well as delays in the database.  In contrast, the rewritten program allows the 
network round trips to be overlapped. It also allows the database to better use 
its resources (multiple CPUs and disks) to process multiple asynchronously 
submitted queries.  Asynchronous calls have been long employed to make 
concurrent use of different system components, like CPU and disk.  

In this 
paper our focus is on automated rewriting of application programs so as to 
submit multiple queries asynchronously, as illustrated in 
\refex{ex:prog2-trans}. In general, automatically transforming a given loop so 
as to make asynchronous query submissions is a non-trivial task, and we 
address the problem in this paper.

The most closely related prior work to our paper is that of Guravannavar and 
Sudarshan~\cite{GUR08}, who describe how to rewrite loops in database 
applications to replace multiple executions of a query in a loop by a single 
execution of a set-oriented (batched) form of the query.  Batching can provide 
significant benefits because it reduces the delay due to multiple synchronous 
round trips to the database, and because it allows more efficient query 
processing techniques to be used at the database. Our program transformation 
techniques for asynchronous query submission are based on the techniques 
described in~\cite{GUR08}, but unlike~\cite{GUR08}, we show how to exploit
asynchronous query submission, instead of batching.

Although batching reduces round-trip delays and allows efficient set-oriented 
execution of queries, it does not overlap client computation with that of the 
server, as the client completely blocks after submitting the batch.  Batching 
also results in a delayed response time, since the initial results from a loop 
appear only after the complete execution of the batch.  Also, batching may not 
be applicable altogether when there is no efficient set-oriented interface for 
the request invoked, as is the case for many Web services.

As compared to batching, asynchronous submission of queries can allow overlap 
of client computation with computation at the server; it can also allow 
initial results to be processed early, instead of waiting for an entire batch 
to be processed at the database, which can lead to better response times for 
initial results.  Further, asynchronous submission is applicable to Web 
Services that do not support set-oriented access. On the other hand pure 
asynchronous submission can lead to higher network overheads, and extra cost 
at the database, as compared to batching.  We present a technique which we call 
asynchronous batching, which combines the benefits of asynchronous submission 
and batching.

The following are the key contributions of this paper:
\begin{enumerate}
 \item We show (in~\refsec{sec:trans}) how a basic set of program 
 transformations, such as loop fission, enable complex programs to be 
 rewritten to make use of asynchronous query submission.  Although loop 
 fission is a well known transformation in compiler optimizations and 
 batching, to the best of our knowledge no prior work shows its use for 
 asynchronous submission of database queries.
 
 \item \refsec{async:sec:sysdesign} describes the design of our 
 implementation.  We first describe (in~\refsec{sec:sysdes-trans}) the design 
 challenges of such a program transformation tool.  Since programmers may need 
 to debug a rewritten version of their program, we present several techniques 
 to make the rewritten program more readable.
 
 We then describe (in~\refsec{sec:sysdes-api}) the design of a framework that 
 supports asynchronous query submission. Our framework provides a common API 
 that can be configured to use either asynchronous submission or batching, or 
 a combination of both.
 
 \item In~\refsec{ext-opt} we present extensions of the basic techniques
 described above.  Specifically, we present (in~\refsec{async-overlap}) a 
 modification of the code generated by the loop fission transformation that 
 optimizes for response time by allowing early generation of initial results.
 
 We also present (in~\refsec{async:sec:basync}) \textit{asynchronous batching},
 a novel technique that combines the benefits of asynchronous query submission 
 and batching by combining, at run time, multiple pending asynchronous 
 requests into one or more batched requests.

 \item These techniques have been incorporated into the DBridge holistic 
 optimization tool~\cite{DBR,SOAP12} to optimize Java programs that use JDBC.
 
 We present (in~\refsec{sec:perf}) a detailed experimental study of the proposed 
 transformations on several real world applications.  The experimental study 
 shows significant performance gains due to our techniques. 
\end{enumerate}

This article is an extended version of our earlier conference 
paper~\cite{CHA11}; the key additions made in this journal version are 
described in~\refsec{sec:relwork}.

The rest of the paper is organized as follows.  A brief background of 
asynchronous submission models is given in~\refsec{sec:setup}. 
Sections~\ref{sec:trans} through~\ref{sec:perf} describe our key 
contributions, as outlined above. Related work is described 
in~\refsec{sec:relwork}.  We discuss possible extensions of our techniques 
in~\refsec{sec:discuss} and conclude in~\refsec{sec:concl}.

\section{Models of Asynchronous Calls} \label{sec:setup}

Two models are prevalent for coordinating asynchronous calls: the
{\em observer model} and the {\em callback model}. \\
\noindent
{\bf The Observer Model:} In this model, the calling program explicitly polls
the status of the asynchronous call it has made. When the results of the call
are strictly necessary to make any further computation, the calling program
blocks until the results are available. The observer model is suitable when the
results of the calls must be processed in the order in which the calls are
made. Example \ref{ex:prog1} of \refsec{sec:intro} shows a program making use of
the observer model to coordinate the asynchronous query execution. We now
formally define the semantics of the methods we use.

\begin{itemize}
\item {\em executeQuery}: Submits a query to the database system
for execution, and returns the results. The call blocks until the query
execution completes.
\item {\em submitQuery}: Submits a query to the database system
for execution, but the call returns immediately with a handle (without
waiting for the query execution to finish).
\item {\em fetchResult}: Given a handle to an already issued query execution
request, this method returns the results of the query. If the query execution
is in progress, this call blocks until the query execution completes. \\
\end{itemize}
\noindent
{\bf The Callback Model:} In this model, the calling program registers a
callback function as part of the non-blocking call. When the request
completes, the callback function is invoked to process the results of the call.
The event driven model is suitable when the program logic to process the call
results is small and the order of processing the results is unimportant.


The program transformations presented in this paper make use of the observer
model for asynchronous query submission. It is possible to extend the proposed
approach to make use of the callback model for programs in which the order of
processing the query result is unimportant. However, the details of such
extensions are not part of this paper.

\section{Basic Transformations} \label{sec:trans} 
Guravannavar et.al.~\cite{GUR08} present a set of program transformation rules
to rewrite program loops so as to enable batched bindings for queries.  In this
section, we show how some of these transformation rules can be extended for
asynchronous query submission.  

The program transformation rules we present, like the equivalence rules of
relational algebra, allow us to repeatedly refine a given program. Applying a
rule to a program involves substituting a program fragment that matches the
antecedent (LHS) of the rule with the program fragment instantiated by the
consequent (RHS) of the rule. Some rules facilitate the application of other
rules and together achieve the goal of replacing a blocking query execution
statement with a non-blocking statement.  Applying any rule results in an
equivalent program and hence the rule application process can be stopped at any
time. We omit a formal proof of correctness for our transformation rules, 
and refer the interested reader to ~\cite{GUR09}.
Each program transformation rule has not only a syntactic pattern to
match, but also certain pre-conditions to be satisfied.  The pre-conditions
make use of the inter-statement data dependencies obtained by static analysis of
the program. Before presenting the formal transformation rules, we briefly
describe the {\em data dependence graph}, which captures the various types
of inter-statement data dependencies.

\subsection{Data Dependence Graph}
\begin{figure}[t]
\begin{minipage} {0.5\textwidth}
\begin{center}
\includegraphics{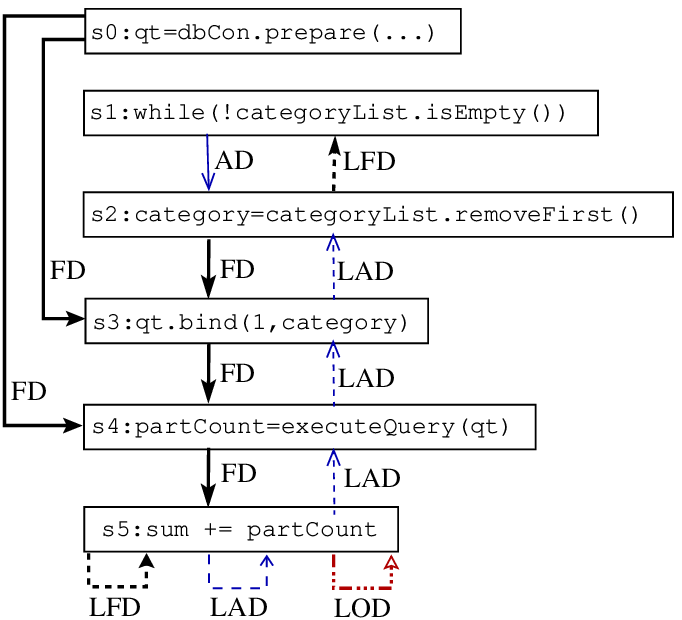}
\caption{Data Dependence Graph for \refex{ex:prog2}}
\label{fig:example-ddg}
\end{center}
\end{minipage}
\end{figure}
Inter-statement dependencies are best represented in the form of a data
dependence graph~\cite{MUCHNICK} or its variant called the program dependence
graph~\cite{PDG87}.  The {\em Data Dependence Graph} (DDG) of a program is a
directed multi-graph in which program statements are nodes, and the
edges represent data dependencies between the statements. The data dependence 
graph for the program of \refex{ex:prog2} is shown in \reffig{fig:example-ddg}.
The types of data dependence edges are explained below.

\begin{itemize}
\item A {\em flow-dependence} edge ($\fdep$) exists from statement (node)
      $s_a$ to statement $s_b$ if $s_a$ writes a location that $s_b$ may
      read, and $s_b$ follows $s_a$ in the forward control-flow. For
      example, in \reffig{fig:example-ddg}, a flow-dependence edge exists
      from node s2 to node s3 because statement s2 writes {\em category}
      and statement s3 reads it.

\item An {\em anti-dependence} edge ($\adep$) exists from statement $s_a$ to
      statement $s_b$ if $s_a$ reads a location that $s_b$ may write, and
      $s_b$ follows $s_a$ in the forward control flow. For
      example, in \reffig{fig:example-ddg}, an anti-dependence edge exists
      from node s1 to node s2 because statement s1 reads {\em categoryList}
      and statement s3 writes it.

\item An {\em output-dependence} edge ($\od$) exists from statement $s_a$ to
      $s_b$ if both $s_a$ and $s_b$ may write to the same location, and $s_b$
      follows $s_a$ in the forward control flow. 

\item A {\em loop-carried flow-dependence} edge ($\lcfd{L}$) exists from
      $s_a$ to $s_b$ if $s_a$ writes a value in some iteration of a
      loop $L$ and $s_b$ may read the value in a later iteration. For 
      example, in \reffig{fig:example-ddg}, 
      a loop-carried flow-dependence edge exists from node s2 to node s1 
      because statement s2 writes {\em categoryList} and statement s1 
      reads it in a subsequent iteration. 
      Similarly, there are loop carried counter parts of {\em anti} and {\em
      output} dependencies, which are denoted by ($\lcad{L}$) and ($\lcod{L}$)
      respectively. 

\item External data dependencies: 
      Program statements may have dependencies not only
      through program variables but also through the database and other
      external resources like files. For example, we have $s_1 \fdep s_2$ if
      $s_1$ writes a value to the database, which $s_2$ may read subsequently.
      Though standard data flow analysis performed by compilers considers only
      dependencies through program variables, it is not hard to extend the
      techniques to consider external dependencies, at least in a conservative
      manner. For instance, we could model the entire database (or file
      system) as a single program variable and thereby assume every query/read
      operation on a database/file to be conflicting with an update/write of
      the database/file. In practice, it is possible to perform a more
      accurate analysis on the external writes and reads.  
      \eat{ 
      When referring to external dependencies explicitly, we use {\em E} as a
      superscript to the corresponding type of dependence edge {\em e.g.,} $s_1
      \extfdep s_2$.
      }
\end{itemize}


\subsection{Basic Loop Fission Transformation}
Consider the program fragment shown in \refex{ex:prog2} and its rewritten
form shown in \refex{ex:prog2-trans}. The key transformation, to enable
such a program rewriting is {\em loop fission} (or {\em loop
distribution})~\cite{KENEDY90}.  Guravannavar et al.~\cite{GUR08} make use of
{\em loop fission} to replace iterative query executions with a batched (or
set-oriented) query execution. In this section, we show how the program
transformation rules proposed in~\cite{GUR08} can be
extended for rewriting programs to make use of asynchronous calls.
\begin{trule}
{\small
\begin{sqlindent}
{\bf while $p$ loop} \\
    \> $ss_1;$ \ s: v = executeQuery(q); \ $ss_2;$ \\
{\bf end loop;} 
\end{sqlindent}
such that:
\begin{enumerate}[(a)]
\item No loop-carried flow dependencies ({\em i.e.,} LCFD edges, external or otherwise) cross the points before and after the query execution statement $s$.
\item No loop-carried {\em external} anti or output dependencies cross the points before and after $s$.
\end{enumerate}

\begin{sqlindent}
\hspace{3cm}{\Large{$\Updownarrow$}} \\
{\em Table(T)} \ t; \\ 
int loopkey = 0; \\
{\bf while $p$  loop} \\
    \> {\em Record(T)} r; \ \  $ss_1'$; \\
    \> r.handle = submitQuery(q); r.key=loopkey++; \\
    \> {\em t.addRecord(r);} \\
{\bf end loop;} 
\\
{\bf for each} r 
               in t {\bf order by} t.key {\bf loop} \\
    \>  $ss_r$; v = fetchResult(r.handle); \ $ss_2$; \\
{\bf end loop;} \\
{\bf delete} t; 
\end{sqlindent}
where the schema $T$ and statement sequences $ss_1'$, $ss_r$ are constructed as follows.\\
Let {\em SV} (split variables) be the set of variables for which 
either an LCAD or LCOD edge crosses the split boundaries (the edge 
is incident from $ss_2$ to $s$ or $ss_1$, or from $s$ to $ss_1$).
\begin{enumerate}
    \item Table $t$ and record $r$ have attributes corresponding to each
          variable in {\em SV} and a key. 

    \item $ss_1'$ is same as $ss_1$ but with additional assignment
          statements to attributes of $r$. Each write to a split variable $v$
          is followed by an assignment statement $r.v = v;$. If the write is
          conditional, then the newly added statement is also conditional on
          the same guard variable.
    \item $ss_r$ is a statement sequence assigning attributes 
          of $r$ to corresponding variables. Each
          assignment in $ss_r$ is conditional; the assignment is made 
          only if the attribute of $r$ is non-null ({\em assigned}).
\end{enumerate}
\caption{Basic Equivalence Rule for Loop Fission}
\label{fig:rule-fission}
}
\end{trule}

A formal specification of the transformation is given as Rule A.
The LHS of the rule is a generic {\em while} loop containing a blocking
query execution statement $s$. $ss_1$ and $ss_2$ are sequences of statements,
which respectively precede and succeed the query execution statement 
in the loop body. The LHS of the rule then lists two pre-conditions, which
are necessary for the rule to be applicable. 
The RHS of the rule contains two loops, the first one making asynchronous
query submissions and the second one performing a blocking fetch followed
by execution of statements that process the query results.

Note that any number of query execution statements within a loop can be
replaced by non-blocking calls by repeatedly applying the loop fission
transformation.  Although we present the loop fission transformation rule
{\em w.r.t.} a {\em while} loop, variants of the same transformation rule can
be used to split set iteration loops (such as the second loop in the {\small
RHS} of the Rule A). 

Rule A makes an improvement of the fundamental nature to the loop fission
transformation proposed in~\cite{GUR08}. Rule A
significantly relaxes the pre-conditions (see Rule 2 in~\cite{GUR08}).
For instance, Rule A allows loop-carried output dependencies to cross
the split boundaries of the loop.  This rule can also be applied to
perform batching, thereby increasing its applicability.  In general,
our transformations are such that the resulting program can be used
either for batching or for asynchronous submission, and this choice can
be made at runtime.  Our transformations in fact blur the distinction 
between batching and asynchronous submission, and can be used to 
achieve the best of both, as described in \refsec{async:sec:basync}.

\subsubsection*{Applicability}
\begin{floatexample}
\begin{sqlindent}
qt = dbCon.prepare(\>\>\>\>``{\bf select} count(partkey) \\
            \>\>\>\>\ \ {\bf from} part {\bf where} p$\_$category=?''); \\ 
category = readInputCategory(); \\
while(category != null) $\{$ \\
\>    qt.bind(1, category); \`(s1) \\
\>    partCount = executeQuery(qt); \`(s2) \\
\>    sum += partCount; \`(s3) \\
\>    category = getParentCategory(category); \`(s4) \\
$\}$ 
\end{sqlindent}
\caption{An example where loop fission is not directly applicable due to loop-carried dependencies}
\label{ex:backdep}
\end{floatexample}

\begin{floatexample}
\begin{sqlindent}
qt = dbCon.prepare(\>\>\>\>``{\bf select} count(partkey) \\
            \>\>\>\>\ \ {\bf from} part {\bf where} p$\_$category=?''); \\
category = readInputCategory(); \\
while(category != null) $\{$ \\
\>    temp$\_$category = category; \\
\>    category = getParentCategory(category);  \\
\>    qt.bind(1, temp$\_$category);  \\
\>    partCount = executeQuery(qt);  \\
\>    sum += partCount;  \\
$\}$ 
\end{sqlindent}
\caption{After reordering the statements in Example \ref{ex:backdep}}
\label{ex:backdep-trans}
\end{floatexample}

The pre-condition that no loop-carried flow dependencies cross the point of
split can limit the applicability of Rule A in several practical
cases.  Consider the program
in \refex{ex:backdep}. We cannot directly split the loop so as to make the
query execution statement (s2) non-blocking, because there are loop-carried
flow-dependencies from statement s4 to s1 and to the loop predicate, which
violate pre-condition (a) of Rule~\ref{fig:rule-fission}.  Statement s4, which
appears after s1, writes a value and statement s1 reads it in a subsequent
iteration. Such cases are very common in practice ({\em e.g.,} in most
{\em while} loops the last statement affects the loop predicate, introducing
a loop-carried flow dependency).

However, in many cases it is possible to reorder the statements within a
loop so as to make loop fission possible, without affecting the correctness of
the program. For example, the statements within the loop of \refex{ex:backdep},
if reordered as shown in \refex{ex:backdep-trans}, permit loop fission.  Note
that in the transformed program of \refex{ex:backdep-trans} there are no
loop-carried flow dependencies, which prohibit the application of 
Rule~\ref{fig:rule-fission} to split the loop at the query execution statement. 
An algorithm for statement reordering to enable loop fission, along with
a sufficient condition for the applicability of the loop fission transformation
are given in \cite{CHA11}.

Further, Rule A is also not directly applicable when
the query execution statement lies inside a compound statement such as an 
\textit{if-then-else} block.  We now present additional transformation rules 
which can be used to address this restriction.

\subsection{Control Dependencies}
\begin{trule}
\begin{transrule}
if $(p)$ $\{$ $ss_1$ $\}$ else $\{$ $ss_2$ $\}$ \\ 
\hspace{10mm} {\large $\Updownarrow$} \\
boolean $cv=p$; \\
{\em ss} \\
\\
where $ss[i]  =  (cv==true)? ss_1[i], 1 \leq i \leq |ss_1|$ and \\
$ss[k+j] = (cv==false)? ss_2[j], 1 \leq j \leq |ss_2|, k=|ss_1|$ 
\end{transrule}
\caption{Converting control-dependencies to flow-dependencies}
\label{rule:control}
\end{trule}

\begin{floatexample}
\begin{sqlindent}
{\bf Initial Program} \\
for (i=0; i $<$ n; i++) $\{$ \\
    \> v = foo(i); \\
    \> if ( v == 0) $\{$ \\
        \>\> v = executeQuery(q); \\
        \>\> log(v); \\
    \> $\}$ \\
    \> print(v); \\
$\}$ \\ \\
{\bf After applying Rule B} \\
for (i = 0; i $<$ n; i++) $\{$ \\
    \> v = foo(i); \\
    \> // {\em Convert control deps to flow deps by}  \\
    \> // {\em making use of a guard variable.} \\
    \> boolean c = (v == 0); \\
    \> c$==$true? v = executeQuery(q); \\
    \> c$==$true? log(v); \\
    \> print(v); \\
$\}$ \\ \\
{\bf After applying Rule A} \\
Table(key, v, c, handle) t; \\
for (i = 0; i $<$ n; i++) $\{$ \\
    \> Record r; \\
    \> v = foo(i); r.v = v; \\
    \> boolean c = (v == 0); r.c = c; \\
    \> c$==$true? r.handle = submitQuery(q); \\
    \> r.key = loopkey++; \\
    \> t.addRecord(r); \\
$\}$ \\
for each r in t order by key loop \\
    \> v = r.v;  c = r.c; handle = r.handle; \\
    \> c$==$true? v = fetchResult(handle);  \\
    \> c$==$true? log(v); \\
    \> print(v); \\
$\}$ 
\end{sqlindent}
\caption{Transforming Control-Dependencies to Flow-Dependencies}
\label{ex:control-to-flow}
\end{floatexample}

We handle control dependencies using the approach of~\cite{GUR08}.
Consider the initial program shown in \refex{ex:control-to-flow}. The query
execution statement appears in a conditional block. This prohibits direct
application of Rule A to split the loop at the program point immediately
following the query execution statement.

Conditional branching ({\em if-then-else}) and {\em while} loops
lead to {\em control dependencies}.  If the predicate evaluated at a conditional
branching statement $s1$ determines whether or not control reaches statement
$s2$, then $s2$ is said to be control dependent on $s1$.  During loop split,
it may be necessary to convert the control dependencies into flow
dependencies~\cite{KENEDY90}, by introducing boolean variables and 
guard statements.  We define a transformation rule to perform this conversion.

The formal specification of the transformation, called Rule 4 in~\cite{GUR08} 
is shown as Rule B in this paper.
An \textit{if-then-else} block is transformed into an assignment of the value 
of the predicate \textit{p} to a boolean variable \textit{cv}, followed by
a sequence of statements guarded by the value (or the negation) of boolean
variable \textit{cv}.
In \refex{ex:control-to-flow}, we apply Rule~\ref{rule:control} and 
introduce a boolean variable $c$ 
to remember the result of the predicate evaluation, and then convert the 
statements inside the conditional block into guarded statements. We can
then apply Rule A and split the loop, as shown in the last part of 
\refex{ex:control-to-flow}.  

\subsection{Nested Loops}
\begin{floatexample}
\begin{sqlindent}
while(pred1) $\{$ \\
\>  while(pred2) $\{$ \\
\>\>    x = executeQuery(q); process(x); \\
\>  $\}$ \\
$\}$ \\
\\
{\bf After Transformation} \\
Table tp; \\
while(pred1)$\{$ \\
\>  Table tc; Record rp; \\
\>  while(pred2)$\{$ \\
\>\>    Record rc; \\
\>\>    rc.handle = submitQuery(q); \\
\>\>    tc.addRecord(rc); \\
\>  $\}$ \\
\>  rp.tc = tc; tp.addRecord(rp); \\
$\}$ \\
for each rp in tp $\{$ \\
\>  for each rc in rp.tc $\{$ \\ 
\>\>    x = fetchResult(rc.handle); process(x); \\
\>  $\}$ \\
$\}$
\end{sqlindent}
\caption{Dealing with nested loops}
\label{ex:nest-loops}
\end{floatexample}

A query execution statement may be present in an inner loop that is nested
within an outer loop. In such a case, it may be possible to split both the
inner and the outer loops, thereby increasing the number of asynchronous query
submissions before a blocking fetch is issued. To achieve this, we first 
split the inner loop and then the outer loop. Such a transformation is
illustrated in \refex{ex:nest-loops}. Note that the temporary table 
introduced during the inner loop's fission becomes a nested table for the 
temporary table introduced during the outer loop's fission. As the idea is 
straight-forward, we omit a formal specification of this rule.

\section{System Design and Implementation} \label{async:sec:sysdesign}
The techniques we propose can be used with any language and data access \api.
We have implemented these ideas and incorporated them into the DBridge holistic
optimization tool~\cite{DBR,SOAP12}.
A system that can support asynchronous query submission would include two
main components (i) a source-to-source program transformer, and (ii) a runtime
asynchronous submission framework.  The runtime infrastructure also supports
asynchronous batching, a technique that supports batching of asynchronous 
requests, described in \refsec{async:sec:basync}.
We now describe each of these components in detail.

\subsection{Program Transformer} \label{sec:sysdes-trans}
Our rewrite rules can conceptually be used with any language. We chose Java as
the target language and {\small JDBC} as the interface for database access.  To
implement the rules we need to perform data flow analysis of the given program
and build the data dependence graph. We used the SOOT optimization
framework~\cite{SOOT}. SOOT uses an intermediate code representation called
Jimple and provides dependency information on Jimple statements. Our
implementation transforms the Jimple code using the dependence information.
Finally, the Jimple code is translated back into a Java program.

\begin{figure}[t]
\begin{minipage} {0.5\textwidth}
\begin{center}
\includegraphics[width=3.4in,height=1.8in]{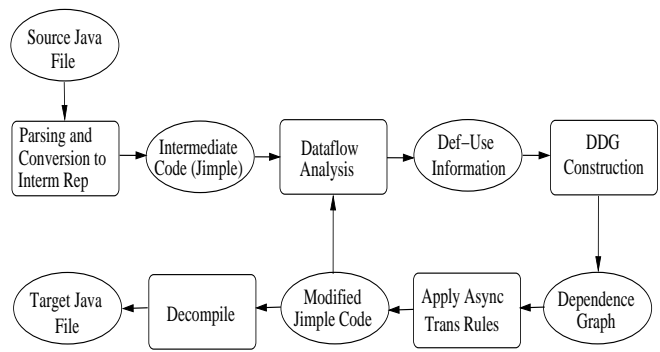}
\caption{Program Transformation Phases}
\label{fig:phases}
\end{center}
\end{minipage}
\end{figure}

The important phases in the program transformation process are shown in
\reffig{fig:phases}. The main task of our program transformation tool appears
in the {\em Apply Async Trans Rules} phase. The program transformation rules
are applied in an iterative manner, updating the data flow information each time
the code changes. The rule application process stops when all (or the user
chosen) query execution statements, which do not lie on a true-dependence
cycle, are converted to asynchronous calls.

Our tool has been implemented with the following design goals.
\begin{enumerate}
\item Readability of the transformed code
\item Robustness for variations in intermediate code
\item Extensibility
\end{enumerate}

Since our program transformations are source-to-source, maintaining readability
of the transformed code is important. We achieve this
goal through several measures. 
(a) The transformed code mostly uses standard JDBC calls
and very few calls to our custom runtime library. This is achieved by providing
a set of JDBC wrapper classes. The JDBC wrapper classes and our custom runtime
library hide the complexity of asynchronous calls. (b) When we apply
Rule~\ref{rule:control} followed by Rule~\ref{fig:rule-fission} to split a
loop, the resulting code will have many guarded statements. This leads to a very
different control structure as compared to the original program. We therefore
introduce a pass where such guarded statements are grouped back in each of the
two generated loops, so that the resulting code resembles the original code.

The intermediate code has the advantage of being simple and suitable for
data-flow analysis, but it makes the task of recognizing desired program
patterns difficult. Each high-level language construct translates to several
instructions in the intermediate representation. We have designed our 
program transformation tool for robust matching of desired program 
fragments. The tool can handle several variations in the intermediate 
(Jimple) code. 

One of our design goals has been extensibility. Each of the transformation
rules has been coded as a separate class. Application of any transformation 
rule independently must preserve the correctness of the program. Such a 
design makes it easy to add new program transformation rules.

\subsection{Runtime Asynchronous Submission Framework} \label{sec:sysdes-api}

The runtime library works as a layer between the actual data access API (such
as JDBC) and the application code. 
It provides asynchronous submission methods in addition to wrapping the 
underlying API. Features such thread management and cache management are 
handled by this library.  The transformed programs in our implementation use 
the \textit{Executor} framework of the \textit{java.util.concurrent} package 
for thread scheduling and management~\cite{EXEC}.  

\begin{figure}[t]
\begin{minipage} {0.5\textwidth}
\begin{center}
\includegraphics[scale=0.18]{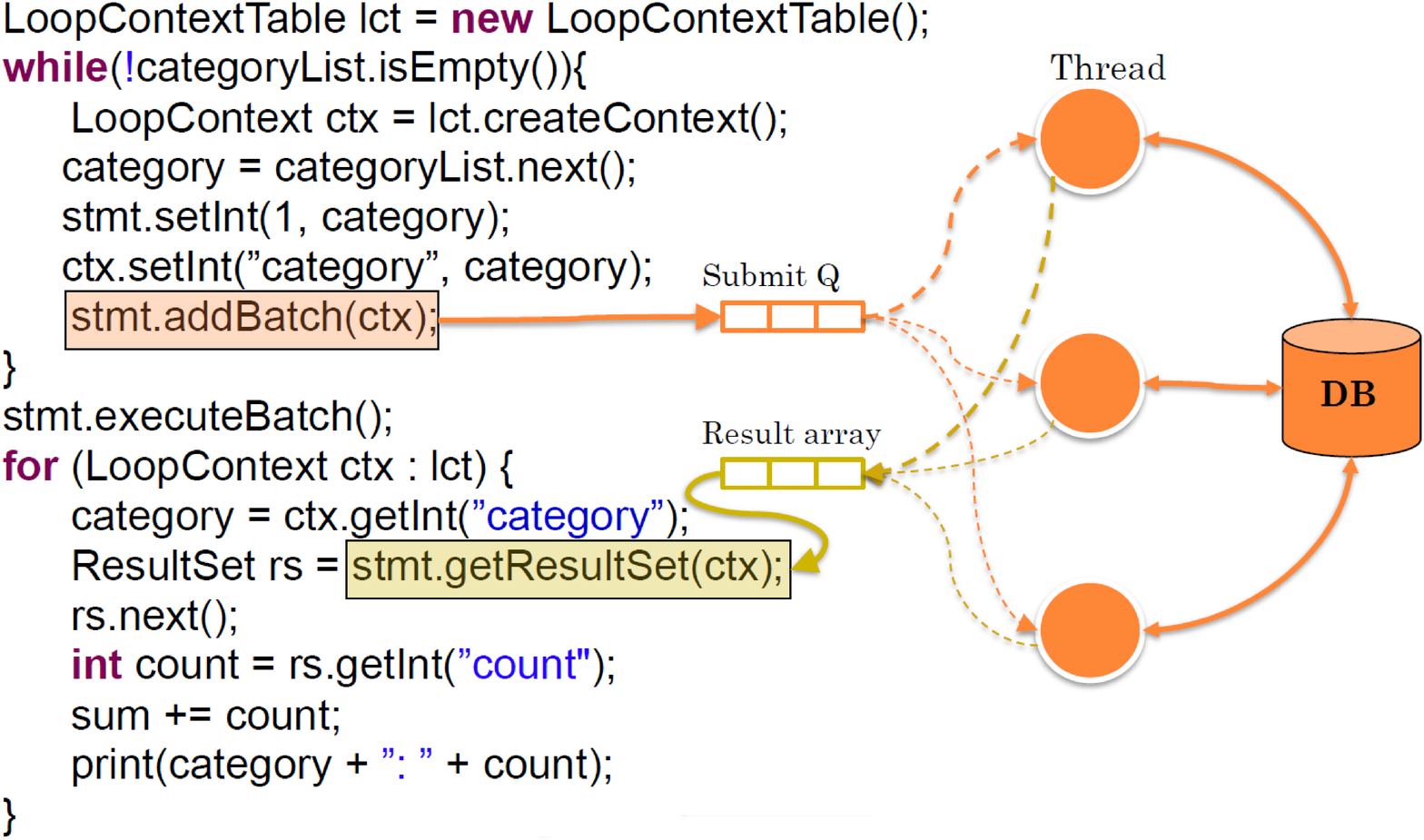}
\caption{Asynchronous query submission API}
\label{fig:async-api}
\end{center}
\end{minipage}
\end{figure}

\reffig{fig:async-api} shows the behaviour of the asynchronous submission \api.
The first loop in the transformed program submits the query to a queue in
every iteration.  The \textit{stmt.addBatch(ctx)} invocation is a non blocking
query submission, with the same semantics as the \textit{submitQuery} \api\ 
described in \refsec{sec:setup}. This queue is monitored by a 
thread pool which manages a configurable number of threads. The requests
are picked up by free threads which maintain open connections to the database. 
The individual threads execute the query in a synchronous manner i.e., the thread 
blocks till the query results are returned.  The results are then placed in 
a cache keyed by the loop context(ctx). 

The second loop accesses the results 
corresponding to the loop context using the \textit{stmt.getResultSet(ctx)} which
has the same semantics as the \textit{fetchResult} \api\ described in 
\refsec{sec:setup}.  Subsequently, it executes statements that depend on 
the query results.  
The \textit{LoopContextTable} ensures the following: (i) it preserves the 
order of execution between the two loops and (ii) for each iteration of the 
first loop, it captures the values of all variables updated, and restores 
those values in the corresponding iteration of the second loop.

\section{Extensions and Optimizations} \label{ext-opt}

We now describe two extensions to our basic technique of asynchronous 
query submission.  These extensions can significantly improve performance
as shown by our experiments.

\subsection{Overlapping the  Generation and Consumption of Asynchronous Requests}
\label{async-overlap}
Consider the basic loop fission transformation Rule A.  A loop when transformed
using this rule, will result in two loops, the first that generates asynchronous
requests (hereafter referred to as the \textit{producer} loop), and the
second that processes, or consumes results (hereafter referred to as
the \textit{consumer} loop).

According to Rule A, the processing of query results (the consumer loop) starts 
only after all asynchronous submissions are completed i.e, after the producer 
loop completes.  Although this transformation significantly reduces 
the total execution time, it results in a situation 
where results start appearing much later than in the original program.  In other 
words, for a loop of \textit{n} iterations, the time to \textit{k}-th response
($1 \leq k \leq n$) for small \textit{k} is more as compared to the original
program, even though the time may be less for larger \textit{k}.  This could be 
a limitation for applications that need to show some results early, or that only
fetch the first few results and discard the rest.

This limitation can be overcome
by overlapping the consumption of query results with the submission 
of requests.  The transformation Rule A can be extended to 
run the producer loop (the loop that makes asynchronous submissions) as a 
separate thread.  That is, the main program spawns a thread to execute
the producer loop, and continues onto the consumer loop immediately.  
Since the loop context table (\textit{Table t} in Rule A) may be empty when 
the consumer loop starts, and may get more tuples as the consumer loop 
progresses, we implement the loop context table as a blocking (producer-consumer) 
queue.  The producer thread submits requests onto this queue, which are picked up 
by the consumer loop.

Note that this transformation is safe, and does not lead to race conditions 
since there are no data dependences between the producer and consumer loop 
other than through the loop context table. This is because the values of all 
variables updated in the producer loop are captured, and restored in the 
consumer loop via the loop context table. The blocking queue implementation of 
the loop context table avoids race conditions on the table. The details of 
this extension are straightforward and hence a formal specification is 
omitted.  We evaluate the benefits of this extension and show the results
in~\refsec{sec:basync-perf}.


\subsection{Asynchronous Submission of Batched queries} \label{async:sec:basync}

As mentioned earlier, the transformation rules proposed in this paper can be 
used for either batching or asynchronous submission.  However, there are key 
differences in the approaches due to which their performance characteristics 
vary.  In this section we compare the relative benefits and drawbacks of 
batching and asynchronous query submission, and propose a new strategy that 
can combine the benefits of both strategies.

\subsubsection{Asynchronous Submission vs. Batching}
%

The following are some of the drawbacks of batching as compared to 
asynchronous submission:
\begin{itemize*}
 \item Although batching reduces round-trip delays and allows efficient 
 set-oriented execution of queries, it does not overlap client computation 
 with that of the server, as the client blocks after submitting the batch.  
 \item  Although batching reduces the overall execution time of the program, 
 for initial results it typically results in a worse response time, since the 
 result of the first query is available only when the result set of the large 
 batch is returned.

 \item Since batching retrieves the results for the whole loop at once, 
 it may significantly increase the memory requirement at the client.
 \item Batching may not be applicable when there is no (efficient)
 set-oriented interface for the request invoked.
\end{itemize*}

The asynchronous query submission technique presented in \refsec{sec:trans} 
avoids the problems mentioned above for batching, but has a few drawbacks of 
its own, as compared to batching:
\begin{itemize*}
 \item Asynchronous query submission does not reduce the number of network 
 round trips but only overlaps them.  This may increase network congestion.
 \item The database still receives individual queries and hence this may 
 result in a lot of random IO at the database.  
 \item As a result of the above, whenever batching is applicable, and the 
 number of iterations of the loop is large, batching leads to much better 
 performance improvements (in terms of total execution time) than asynchronous 
 submission.
\end{itemize*}
More details are described in our experiments in~\refsec{sec:basync-perf}.  We 
now describe how to combine both these approaches.

\subsubsection{Asynchronous Batching: Best of Both Worlds} \label{basync-main}
Batching and Asynchronous submission can be seen as two ends of a spectrum.
Batching, at one end, combines all requests into one big request with no 
overlapping
execution, where as asynchronous submission retains individual requests as is, 
while completely overlapping their execution.  Clearly, there is a range of 
possibilities between these two, that can be achieved by \textit{asynchronous}
submission of multiple, smaller \textit{batches} of queries.  This approach, 
which we call \textit{asynchronous batching}, retains the advantages 
of batching and asynchronous submission, while avoiding their drawbacks.

Consider the example in \reffig{fig:async-api}.  As
mentioned earlier, the first loop in this program submits a query to a queue 
in each iteration.  This request queue is monitored by a thread pool. 
In pure asynchronous submission, each free thread picks up an individual 
request from the queue.  In contrast, with asynchronous batching, the thread 
can observe the whole queue, and pick up one, or more, or all requests from 
the queue.  
If a thread picks up a single request, it executes the query as described 
earlier. However. if a thread picks up more than one request, it performs a 
query rewrite as done in batching, and executes those requests as a batch.  
Once the result of the batch arrives, it is split into multiple result sets 
corresponding to each individual query, which are then placed in the cache.  

Asynchronous batching aims to achieve the best of batching and asynchronous
submission, since it has the following characteristics.
\begin{itemize*}
 \item Like batching, it reduces network round trips, since multiple requests 
 may be batched together.
 \item Like asynchronous submission, it overlaps client computation with that 
 of the server, since batches are submitted asynchronously.
 \item Like batching, it reduces random IO at the database, due to use of set 
 oriented plans. 
 \item Although the total execution time of this approach might be comparable
 to that of batching, this approach results in a much better response time 
 comparable to asynchronous submission, since the results of queries become 
 available much earlier than in batching.
 \item Memory requirements do not grow as much as with pure batching, since we 
 deal with smaller batches.
\end{itemize*}

The key challenge in engineering such a system is to identify the sweet spot 
in the spectrum between batching and asynchronous submission.  This primarily 
involves deciding the size of each batch and 
the number of threads to use, which would result in the best performance.  
This decision cannot be made statically during program transformation, since
it depends on runtime factors such as (i) the number of iterations in the loop,
(ii) the query processing time and the size of its results, (iii) the 
capacity and load on the client machine and the database server, (iv) network 
bandwidth availability.

Asynchronous batching is a completely runtime decision; the program 
transformation is performed in accordance with the rewrite rules in this 
paper, and requires no additional rewriting. 
The runtime library makes decisions on asynchronous calls vs. partial batching
in a dynamic fashion.  We now 
discuss strategies to tune parameters for asynchronous batching.

\subsubsection{Adaptive tuning of parameters} \label{tuning}

The runtime library is extended to allow a thread to pick up one or more 
requests from the queue.  However the key problem is the following: given a
queue of $n$ requests, how many requests should a free thread pick up?
Note that the only information available for a thread is the current state of
the queue.
In order to simplify our discussion, we make a few performance related assumptions:
\begin{itemize*}
 \item The database is not under heavy load and is able to handle 
 concurrent requests efficiently.
 \item The number of threads $T$ available on the client is fixed.
 \item The network characteristics do not vary drastically during program 
 execution.
\end{itemize*}
Given these assumptions, we now propose strategies to automatically vary the 
batch size at runtime.  These strategies are affected by the following metrics:

\begin{enumerate}
 \item \textit{The request arrival rate:}
  This is the rate at which the 
program submits requests onto the queue.  In our example of \reffig{fig:async-api},
the first loop submits one request per iteration.  So this arrival rate 
essentially captures the time taken by each iteration of the first loop before
submitting a request.  If there are expensive operations in this loop such as
remote calls, they affect the request arrival rate.
 \item \textit{The request processing rate:} This is the rate at which requests
 in the queue are processed. 
Processing a request includes the query processing time
at the database and the network round trip time.  Since we only consider cases
where the query is the same in each iteration, with varying parameter values,
we assume that the query processing time is the same for each request.
\end{enumerate}
The request arrival rate would be higher than the requests processing rate
if any of the following are true: 
(a) the producer loop has no expensive operations,
(b) network round trips are very expensive, 
(c) query processing time is high.
We now propose three possible strategies for asynchronous batching.

~\\
\noindent\textbf{One-or-all Strategy:} 
This is a simple strategy to combine asynchronous submission and batching.
Given a queue with $n$ requests, the \textit{One-or-all} strategy for a free 
thread 
is as follows: If $n=1$, then pick up the request from the queue, and execute 
it as an individual request. If $n>1$, pick up all the $n$ requests in the 
queue and batch them.  In other words, (i) insert the parameters of the $n$ 
requests
into a temporary parameter table, (ii) rewrite the query using the technique 
given in \cite{GUR08}, (iii) execute this rewritten query.  If $n = 0$, wait 
for new requests.  In this strategy, a free thread will always clear the queue 
by picking up all pending requests from the queue.  

~\\
\noindent\textbf{Lower Threshold Strategy:} 
The \textit{One-or-all} strategy can be improved 
based on an observation regarding 
batching.  Batching results in 3 network round trips, one each for 
(a) inserting parameters into a temporary table, 
(b) executing the batched query, and 
(c) clearing the temporary table.  
In fact each thread incurs another round trip while batching for the first 
time in order to create the temporary table. 
This means that the time taken to process one batch is roughly equivalent to 
the time taken to process \textit{at least} three individual 
requests sequentially, since there are 3 network round trips and 3 queries 
being executed for every batch.  We verified this in our experiments, and 
found that very small batches perform poorly as compared to 
asynchronous submission.  

Therefore, we use the following strategy.
We define a \textit{batching threshold} $bt \ge 3$. 
If $n > bt$, then pick up all the $n$ requests in the queue and batch them.  If
$1 \le n \le bt$, then pick up one request from the queue, and execute 
it as an individual request. If $n = 0$, wait for new requests.  Observe that
in this strategy, a free thread does not necessarily clear the queue.

Consider the situation where the request arrival rate is higher than the 
request processing rate. 
In this setting, the first few (about $T$) requests would be sent as 
individual requests asynchronously.  Since the 
queue builds up much faster than it is consumed, after the first few 
iterations, the requests would be submitted in batches with increasing sizes.  

On the other hand, consider the case where the request processing rate is 
higher than the rate of arrival of requests onto the queue.  In this situation, 
the queue would not grow in size since the requests keep getting consumed at a 
higher rate, and hence \textit{n} would remain below (or close to) the 
\textit{batching threshold}.  This implies that most requests would be sent
individually, mimicking the behaviour of asynchronous query submission.  

Thus 
we can see that the lower threshold strategy is actually quite adaptive. 
Batch sizes vary in accordance with the queue size, which in turn depends upon 
the arrival rate of requests, the rate at which requests get processed, and
the number of threads working concurrently on processing requests.

~\\
\noindent\textbf{Growing upper-threshold based Strategy:} 
Although the above approach improves response time and adapts the batch size
according to the queue size, in situations where the arrival rate of requests 
is high, it may lead to a situation where a single large batch is submitted 
while the remaining threads are idle.  This could lead to a slower response 
time for initial results, since the database would take a longer time to 
process a large batch, and higher memory consumption due to a large request
queue, although the larger batch size may reduce overall work at the database 
server, and reduce the time to process all requests.

For applications that need better response times for initial results, we use 
an upper-threshold strategy.  We use a 
\textit{growing upper threshold} that bounds the maximum batch size. This 
upper threshold is not a constant but is initially small, so that batch sizes 
are small initially, but grows as more requests are submitted, so that 
response times for later results are not unduly affected due to very small 
batch sizes.

The growing upper-threshold strategy works as follows. If the number of 
requests in the queue is less than the current upper threshold, all requests 
in the queue are added to a single batch. However, if the number of requests 
in the queue is more than the current upper threshold, the batch size that
is generated is equal to the current threshold; however, for future batches, 
the upper threshold is increased; in our current implementation of the growing 
upper-threshold strategy, we double the upper threshold whenever a batch of 
size equal to the current upper threshold is created.  

Note that the upper 
threshold strategy is orthogonal to the lower-threshold strategy, and each may 
be used with or without the other.

\section{Experimental Results} \label{sec:perf}
We have conducted a detailed experimental evaluation of our techniques using
the DBridge tool.  In \refsec{subsec:asyncperf}, we present our experiments 
on asynchronous query submission and its benefits. Next, in 
\refsec{sec:basync-perf}, we compare basic asynchronous submission with the 
extensions and optimizations described in \refsec{ext-opt}, and discuss the 
results.  

\subsection{Asynchronous query submission} \label{subsec:asyncperf}
For evaluating the applicability and benefits of the proposed transformations,
we consider four Java applications: two publicly available benchmarks
(which were also considered by Manjhi et.al.~\cite{AMIT09}) and two other
real-world applications we encountered. Our current implementation does not
support all the transformation rules presented in this paper, and does not
support exception handling code. Hence, in some
cases part of the rewriting was performed manually in accordance with the
transformation rules. 

We performed the experiments with two widely used
database systems - a commercial system we call {\small SYS1}, and PostgreSQL.
The SYS1 database server was running on a 64 bit dual-core machine with 4 GB 
of RAM, and PostgreSQL was running on a machine with two Xeon 3 GHz 
processors and 4 GB of RAM. 
Since disk IO is an important parameter that affects the performance of
applications, we report the results for both warm cache and cold cache. The
Java applications were run from a remote machine connected to the database servers
over a 100 Mbps LAN. The applications used JDBC API for database connectivity.
The cache of results was maintained using the \textit{ehcache} 
library~\cite{EHCACHE}.

\begin{figure}[t]
\begin{center}
 \includegraphics[scale=.3, angle=270]{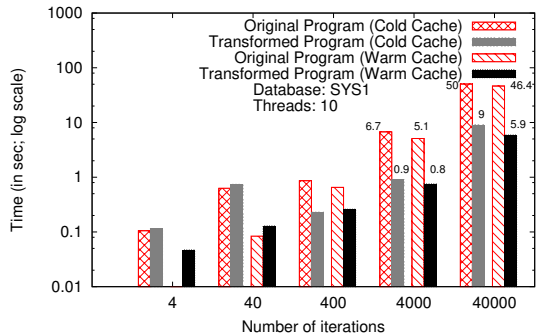}
 \caption{Experiment 1 with varying number of iterations}
 \label{fig:rubis-iter}
\end{center}
\vspace{-5mm}
\end{figure}

\mysubsection{Experiment 1: Auction Application} 
We consider a benchmark application called RUBiS~\cite{JMOB} that represents a
real world auction system modeled after \textit{ebay.com}. The application
has a loop that iterates over a collection of comments, and for each
comment loads the information about the author of the comment. The 
\textit{comments} table had close to 600,000 rows, and the \textit{users}
table had 1 million rows.  First, we consider the impact of our transformations
as we vary the number of loop iterations (by choosing user ids with appropriate 
number of associated comments), fixing the number of threads at 10.
\reffig{fig:rubis-iter} shows the performance of this program before and after
the transformations with warm and cold caches in log scale. The y-axis denotes
the end to end time taken for the loop to execute, which includes the
application time and the query execution time.

For a small number of iterations, the transformed program is slower than the 
original program. The overhead of thread creation and scheduling overshoots 
the query execution time. However, as the number of iterations increases, the 
benefits of our transformations increase. For the case of 40,000 iterations, 
we see an improvement of a factor of 8.

\begin{figure}[t]
\begin{center}
 \includegraphics[scale=.3, angle=270]{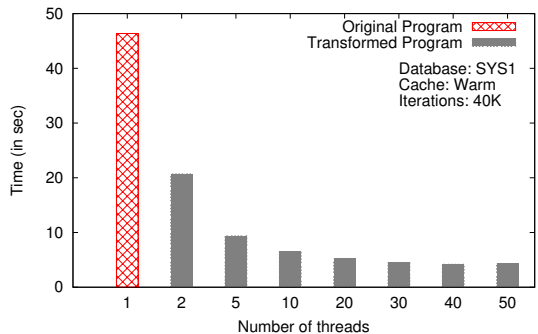}
 \caption{Experiment 1 with varying number of threads }
 \label{fig:rubis-thread}
\end{center}
\vspace{-5mm}
\end{figure}

Next, we keep the number of iterations constant (at 40,000) and vary the number 
of threads. The results of this experiment are shown in \reffig{fig:rubis-thread}. 
The execution time (for both the warm and cold cache) drops sharply as 
the number of threads is increased, but gradually reaches a point where the 
addition of threads does not improve the execution time. 
The results of the above experiment on PostgreSQL follow the same pattern
as in the case of SYS1, and the results are given in~\cite{CHA11}.

\begin{figure}[t]
\begin{center}
 \includegraphics[scale=.3, angle=270]{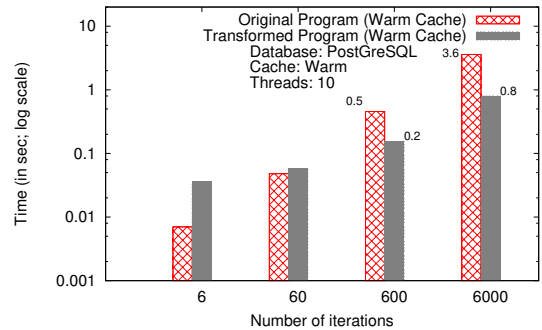}
 \caption{Experiment 2 with varying number of iterations}
 \label{fig:bboard-iter-postgres}
\end{center}
\vspace{-5mm}
\end{figure}

\mysubsection{Experiment 2: Bulletin Board Application} 
RUBBoS~\cite{JMOB} is a benchmark bulletin board-like system inspired by \textit{slashdot.org}.
For our experiments we consider the scenario of listing the top stories of the day, along 
with details of the users who posted them. \reffig{fig:bboard-iter-postgres} shows the results
of our transformations with different number of iterations. Although the transformed program 
takes slightly longer time for small number of iterations, the benefits increase with the number
of iterations (note the log scale of y-axis).

\mysubsection{ Experiment 3: Category Traversal} 
This program, taken from ~\cite{GUR08},  finds the part with maximum
size under a given category (including all its sub-categories) by performing a
DFS of the category hierarchy. For each node (category) visited, the program
queries the item table. The TPC-H part table, augmented with a new column
category-id and populated with 10 million rows, was used as the item table. The
category table had 1000 rows - 900 leaf level, 90 middle level and 10 top level
categories (approximately). A clustering index was present on the category-id
column of the category table and a secondary index was present on the
category-id column of the item table. 

\begin{figure}[t]
\begin{center}
 \includegraphics[scale=.3, angle=270]{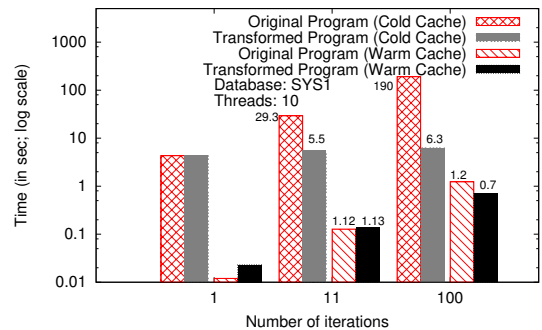}
 \caption{Experiment 3 with varying iterations}
 \label{fig:travcat-iter}
\end{center}
\vspace{-5mm}
\end{figure}

\reffig{fig:travcat-iter} shows the performance of this program before and
after applying our transformation rules. As in the earlier example, we first
fix the number of threads and vary the number of iterations. We perform this
experiment with ten threads, on a warm cache on SYS1. The results are in accordance with our
earlier experiments. In addition, we observe that the  number of threads is an
important parameter in such scenarios.  This parameter is influenced by several
factors, such as the number of processor cores available for the database
server and the client, the load on the database server, the amount of disk IO,
CPU utilization etc. 

When the program is
run with a cold cache, the amount of disk IO involved in running the queries is substantially
higher than with a warm cache. But the bottleneck of disk IO can be reduced by issuing 
overlapping requests. Such overlapping query submissions enable the database system to 
choose plan strategies such as shared scan. 

%
The effect of varying the number of threads shows similar trends as that
of Experiment 1, though the actual numbers differ.  The results can be found 
in \cite{CHA11}.
In transforming this program, the reordering algorithm was first applied and then the loop was 
split using Rule A.

\mysubsection{Experiment 4: Web service invocation} 
Although we presented our program transformation techniques in the context of
database queries, the techniques are more general in their applicability, and 
can be used with requests such as Web service calls. 
In this experiment, we consider an application that fetches data about directors and 
their movies from Freebase~\cite{FREEBASE}, a social database about entities, spanning 
millions of topics in thousands of categories. It is an entity graph which 
can be traversed using an API built using JSON over HTTP. The client application, 
written in Java, retrieves the movie and
actor information for all actors associated with a director. Such applications
usually require the execution of a sequence of queries from within a loop because 
(a) operations such as joins are not possible directly, and (b) the Web service
API may not support set oriented queries. 

Since our current implementation supports only JDBC API, we manually applied
the transformations to the code that executes the Web service requests.
\begin{figure}[t]
\begin{center}
 \includegraphics[scale=.3, angle=270]{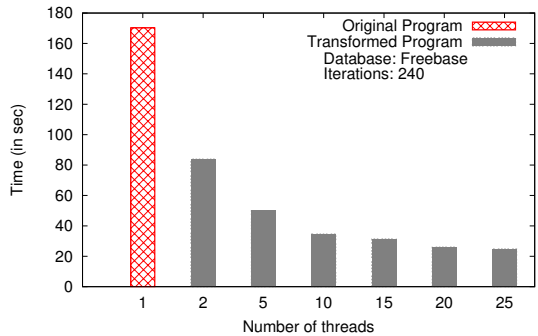}
 \caption{Experiment 4 with varying number of threads}
 \label{fig:movie-thread}
\end{center}
\end{figure}
The results of this experiment are shown in \reffig{fig:movie-thread}. As we
vary the number of threads, overlapping HTTP requests are made by the client
application which saves on network round-trip delays. Since our experiment used
the publicly available Freebase sandbox over the Internet, the actual time
taken can vary with network load. However, we expect the relative improvement
of the transformed program to remain the same.  

\mysubsection{Time Taken for Program Transformation}
Although the time taken for program transformation is usually not a concern 
(as it is a one-time activity), we note that, in our experiments the 
transformation took very little time (less than a second).

\subsection{Applicability of Transformation Rules} 

\begin{table}[t]
\caption{Applicability of transformation rules} \label{tab:applicability}
	\begin{tabular}{ | l | l | l | l | } 
		\hline 
		Application & \# Opportunities & \# Transformed & Applicability (\%)\\ \hline 
		Auction & 9 & 9 & 100\\ \hline 
		Bulletin Board & 8 & 6 & 75 \\ \hline 
	\end{tabular}
\end{table}

In order to evaluate the applicability of our transformation rules, we
consider the two publicly available benchmark applications used above, the auction
application and the bulletin board application. For each of these, 
we have analyzed the source code to find out (a) how many opportunities for
asynchronous submission of queries exist, and (b) how many of those
opportunities are exploited by our transformation rules. The results of
the analysis is presented in Table \ref{tab:applicability}.

We consider all kinds of loop structures which include a query execution
statement in the loop body, as potential opportunities (\# Opportunities). 
Among such 
potential opportunities, those which satisfy the preconditions for our
rules, are exploited (\# Transformed). This would involve reordering of statements in a 
lot of situations. 

We see that all such opportunities present in the auction system indeed
satisfy the preconditions and can be transformed.  Although our preconditions are 
more general than those proposed in~\cite{GUR08}, the opportunities satisfied both.
In the bulletin board
application, few of the loops performed recursive method invocations
which prevent them from being transformed. 
Out of the programs seen earlier, the remaining were too small for this 
analysis, and hence omitted.

\subsection{Effect of Optimizations} \label{sec:basync-perf}

\begin{figure}[t]
\begin{center}
 \includegraphics[scale=.3, angle=270]{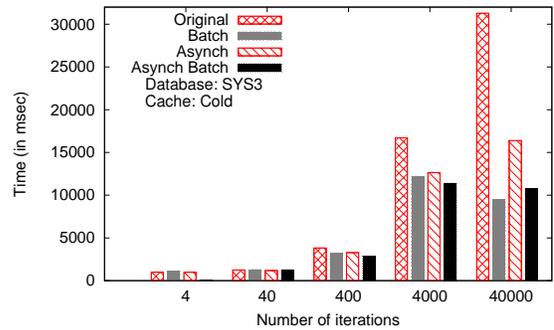}
 \caption{Total execution time with no. of iterations}
 \label{fig:vui-basync}
\end{center}
\vspace{-5mm}
\end{figure}

\begin{figure}[t]
\begin{center}
 \includegraphics[scale=.6, angle=270]{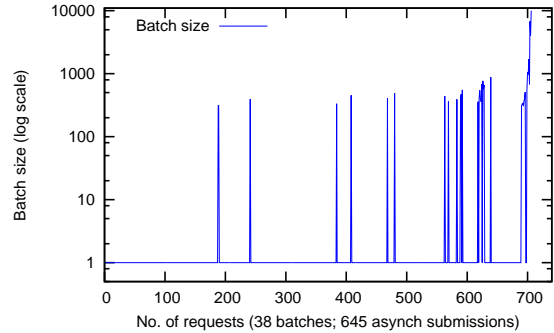}
 \caption{Batch sizes during an execution}
 \label{fig:vui-basync-beh}
\end{center}
\vspace{-5mm}
\end{figure}

\begin{figure}[t]
\begin{center}
 \includegraphics[scale=.72, angle=270]{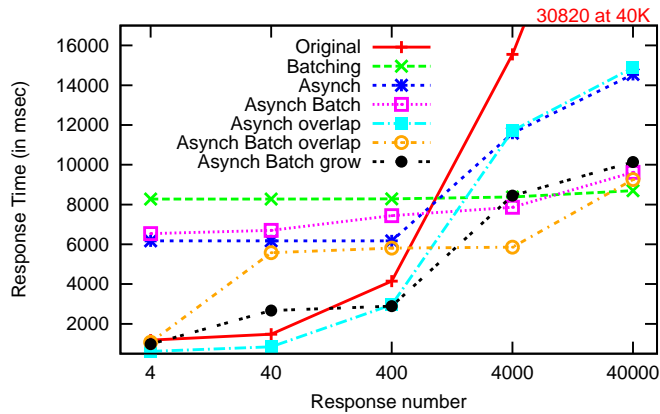}
 \caption{Time to \textit{k}-th response (with 40000 iterations)}
 \label{fig:vui-basync-resp}
\end{center}
\vspace{-5mm}
\end{figure}

We have performed experiments to compare the following approaches:
(i) \textit{Original}: the original program 
(ii) \textit{Batch}: the program rewritten using query batching, 
(iii) \textit{Asynch}: the program rewritten according to our technique 
of asynchronous submission, 
(iv) \textit{Asynch Batch}: our technique of combining batching and
asynchronous submission (\refsec{async:sec:basync}), using the simple 
threshold based strategy
(v) \textit{Asynch Overlap}: asynchronous submission with concurrent 
generation of requests (\refsec{async-overlap}),
(vi) \textit{Asynch Batch Overlap}: asynchronous batching with concurrent 
generation of requests, and 
(vii) \textit{Asynch Batch Grow}: asynchronous batching with concurrent 
generation of requests and the growing-upper-threshold strategy.
Our current implementation does not support the \textit{Async Overlap}
transformation, and hence we have rewritten the code manually
as described in \refsec{async-overlap}.

The experiments have been conducted on a widely used commercial database 
system {\small SYS3}. The {\small SYS3} database server was running on a 64 
bit 2.3 GHz quad-core machine with 4 GB of RAM.  The Java applications were 
run from a remote 3.3 Ghz quad-core machine connected to the database servers 
over a 100 MBps LAN.  
For approach (iv), we used a lower batching threshold of 300 with 48 threads, and 
for approach (vii), we used a doubling growth rate for the upper threshold. 
We again consider the benchmark auction application 
RUBiS~\cite{JMOB} and the scenario described in Experiment 1 of 
\refsec{subsec:asyncperf}.

\subsubsection{Total execution time}
First, we compare the total execution time of this program according to the 
approaches (i) through (iv), since the optimizations in (v), (vi) and (vii)
have minimal impact on the total execution time.
The results of this experiment with cold cache are shown in 
\reffig{fig:vui-basync}.  The x-axis shows the number of iterations and the 
y-axis shows the total execution time in milliseconds.  

It can be observed from \reffig{fig:vui-basync} that at smaller number of 
iterations, all approaches 
behave very similarly, and differences can be observed at larger number of 
iterations.  Asynchronous submission (with 12 threads) gives about 50\% 
improvement, while batching leads to about 75\% at 40000 iterations.  
Asynchronous batching, with 48 threads and a lower batching threshold of 300 
leads to about 70\% improvement. 

At 40000 iterations, we have recorded the behaviour of one run of asynchronous 
batching, shown in \reffig{fig:vui-basync-beh}.  The x-axis shows the number 
of requests (either batched or individual), and the y-axis shows the batch 
sizes in log scale.  Overall, there were 38 batch submissions and 645 
asynchronous submissions, and among the 38 batches, the average batch size was 
1019.  Initially, many requests are sent individually since the lower batching 
threshold was set to 300.  But the queue builds up quite fast and hence there 
are a few intermittent batch submissions.  As the execution progresses, there 
are more and more batch submissions, and batch sizes also start growing.  
Towards the end, there are batches of upto 10000 requests.  This behaviour is 
in accordance with our expectation as described in \refsec{async:sec:basync}.

\subsubsection{Time to \textit{k}-th response}
Next, we compare the response time of the program according to approaches
(i) through (vii) described earlier.  Here, by response time we mean
the duration between the start of the program and the arrival (or the output)
of the \textit{k}-th response from the program.  In our auction system 
experiment, records are printed when the information about the author of the 
comment is retrieved.  Therefore, the response time is measured at the instant 
where the author information of the \textit{k}-th comment is output.
We fix the number of iterations at 40000, and record the time taken for the
\textit{k}-th response, with \textit{k} varying from 1 to 40000.  The results 
of this experiment are shown in \reffig{fig:vui-basync-resp}.  The x-axis 
shows the response number \textit{k}, and the y-axis shows time in 
milliseconds.  For this experiment, the
\textit{Async Batch Grow} approach used a lower batching threshold
of 100, and an upper threshold that doubles, initially set to
200.

The original program has the best response time initially.
However, the response time increases quite steeply with \textit{k}, and 
reaches about 31 seconds for the 40000th response.
Batching, in contrast, has a constant curve.  This is because
even the first response is output only after (i) all parameters are
added to the parameter batch table, and (ii) the transformed (set oriented) query is 
executed.  Essentially, the time to \textit{k}-th response in batching is very 
close to the total execution time, since all the results are returned together.

The response time for \textit{Asynch} starts off with a better (lower) 
response time as compared to batching, but increases beyond batching for 
larger values of \textit{k}.  Asynchronous batching, initially
behaves similar to asynchronous submission, and slowly deviates from it.  At larger
number of iterations, it behaves more similar to batching.  In other words,
it always tends towards the better of \textit{Asynch} and \textit{Batch}.

The \textit{Overlap} versions of \textit{Asynch} and \textit{Asynch Batch}
show much better response times compared to the earlier approaches. 
The \textit{Async Batch Grow} approach behaves the best in balancing response
time vs total execution time.  It initially shows response times similar to
the original program, and does even better than the \textit{asynch} and 
\textit{Batch} at larger iterations.  At $k=40000$, it results in the response
time comparable to \textit{Batch}.

\subsubsection{Discussion}
In summary, our experimental study shows that batching and asynchronous submission
are beneficial techniques with different trade offs, and the combined technique
of asynchronous batching with optimizations aims at balancing these trade offs.  
Some of the trade offs are (a) total execution time vs. time to 
\textit{k}-th response, (b) reducing network round trips (by batching multiple 
requests) vs. overlapping execution of queries, (c) reducing memory 
consumption (by using iterative query execution) vs. set oriented execution of 
the query.

These trade offs are essentially controlled by the parameters used in 
asynchronous batching, such as the batching threshold, number of threads etc.
Based on the use case, the parameters have to be tuned in order to achieve
the desired behaviour.  Our contribution in this paper has been to expose 
these trade offs to the developer, and allow manual tuning of such parameters.  
We have also presented some initial approaches for automatic tuning of 
parameters.  Although our approaches are quite adaptive, as described in 
\refsec{tuning}, we believe that there is scope for more work in this area.

\section{Related Work} \label{sec:relwork}

Most operating systems
today allow applications to issue asynchronous IO requests~\cite{LINUXAIO}.
Asynchronous calls are also used for data prefetch and overlapping operator
execution inside query execution engines ~\cite{GRA03, GAL07, IYENGAR08}.
Asynchronous calls have also been used to hide memory access latency by issuing
prefetch requests~\cite{STEVEN00}.  Asynchronous calls are widely used in 
the communication between the web browser and the server using manually 
placed AJAX requests.


Yeung~\cite{KWOK04} proposes an approach to automatically optimize distributed
applications written using Java RMI based on the concept of deferred execution, where
remote calls are delayed for as long as possible.  Such delaying 
enables optimizations such as call aggregation, server
forwarding etc.  However, this work does not consider asynchronous calls and 
and query executions within loops.


Dasgupta et al.~\cite{Arjun09} and Chaudhuri et 
al.~\cite{Surajit09} propose an architecture and techniques for a general 
static analysis framework to analyze database application binaries that use the 
{\small ADO.NET API}, with the goals of identifying security, correctness
and performance problems.  
Like these approaches, we too use static analysis, but specifically for
optimization by introducing asynchronous prefetching of query results.

There has been work on prediction based prefetching of query results 
\cite{BowmanS07}, by analyzing logs and trace files, but this work does not
consider asynchronous prefetching.  
There has been very recent work on automatic partitioning of database 
applications by Cheung et al.~\cite{statusquo}, with the goal of eliminating 
the many small, latency-inducing round trips between the application and 
database servers. 
However, their approach does not exploit the opportunities that
arise due to program transformations, and overlapping of computation by
asynchronous submission of queries.


Guravannavar et.al.~\cite{GUR08} consider rewriting loops in database
applications and stored procedures, to transform iterative executions of
queries into a single execution of a set-oriented form of
the query. We use a similar framework of program transformation, but
for asynchronous query submission. 

While our transformation rules are based on \cite{GUR08}, we 
make the following novel contributions.  First, we show how the transformation 
rules presented in \cite{GUR08} in the context of batching, can be
adapted for asynchronous query submission.
Second, we describe an extension to our transformation that enables overlapping 
of generation and consumption of asynchronous requests, thereby greatly improving 
the response time.  Third, we present a technique to combine batching and 
asynchronous query submission into a common framework. Also, we describe an 
infrastructure to support asynchronous query submission, and the challenges and 
trade offs in designing and implementing such infrastructure.



Manjhi et al.~\cite{AMIT08} consider prefetching of query results by employing 
non-blocking database calls, made at 
the beginning of a function. A blocking call is subsequently issued when the 
results of the query are needed, and this call is likely to take much less time 
as the query results would be already computed and available in the cache.
However, they do not describe details to automate this task, and also
do not consider loops and procedure invocations. 

Ramachandra et al.~\cite{RAM12} propose a technique to insert prefetch requests
for queries/web services at the earliest possible point in the program across 
procedure invocations.  However, they do not consider loop
transformations for queries within loops while exploiting opportunities for 
prefetching, and this forms the main focus of this paper.  

The approaches described in \cite{AMIT08, RAM12} integrate well with our 
technique of transforming programs to enable asynchronous submission of 
queries.  Consider cases where a loop invokes a procedure which in turn 
executes a query.  Such cases are quite common in applications backed
by object relational mappers such as Hibernate~\cite{HIB}.  They can be 
optimized by first applying the prefetching technique described in \cite{RAM12}
which brings the prefetch instruction directly into the loop.  Subsequently,
the loop transformations presented in this paper can be applied.


\textit{The present article is an extended version of the conference paper~\cite{CHA11},
with the following key differences}.  
The extensions presented in \refsec{ext-opt}, and the related performance 
experiments in \refsec{sec:basync-perf} are entirely novel, as are most of the 
system design issues in \refsec{async:sec:sysdesign}, and the discussions in 
\refsec{sec:discuss}.  Due to lack of space, we have omitted details of statement
reordering to improve the applicability of our transformations, and a few of the 
experimental results from~\cite{CHA11}.

%
%

\section{Extensions} \label{sec:discuss}
We now discuss some system design considerations and extensions
of techniques described in this paper.

\noindent{\bf Ensuring transaction properties}:
In our implementation, we have used one connection per thread in order to 
achieve overlapping query execution.  This is because in JDBC, (a) a database 
connection allows only one open query at a time, (b) there are no API methods
that allow asynchronous submission.  ADO.NET provides asynchronous API (such 
as the \textit{BeginExecuteReader} and \textit{EndExecuteReader} APIs), which
allow overlapping of query execution with local computation.  However, even 
these APIs do not support overlapping query executions through a single connection.

In order to fully preserve transaction properties and achieve 
true asynchronous submission, individual threads in the thread pool 
should be part of a single shared transaction.  Such an infrastructure
is not currently supported by any database vendor to the best of our knowledge.
Although databases support distributed transactions (such as JDBC XA 
transactions), their goal is to allow transactions across multiple data sources.

One way to implement this (if snapshot queries are supported) is to allow 
multiple connections to share a snapshot 
point.  Such a feature, if supported, would allow multiple threads (with their own connections) to 
share and execute transactions on the same snapshot.  We believe that
this would be a minor change in databases that already support snapshot 
isolation, and would be a useful feature to have.  Such a built in support
would not only simplify application development, but also lead to significant
improvement in performance, as compared to our current implementation.

Rewriting loops containing update transactions needs to consider dependencies
between update statements and program variables.  A conservative approach
is to assume that update statements are dependent on other update or select
statements in a loop, and model them as data dependencies which factor in to
the preconditions for our transformation rules.  This can be improved by using 
more precise inter query dependence analyses \cite{Par96}.

\noindent{\bf Minimizing memory overheads}: 
If the number of loop iterations is large, the transformed program may incur high
memory overhead, in order to store the handle and the state associated with each 
iteration.  Storing such state on disk increases the IO cost.  Our technique can be 
extended such that, based on memory usage, the producer thread backs off
and waits while results are consumed and memory freed, and then generates more 
requests.


\noindent{\bf Which calls to be transformed?}: 
It may not be beneficial to transform every blocking query submission call to a
non-blocking call. From our experimental study it is also evident that given a
query execution statement, the benefit to be achieved by converting it to a
non-blocking call depends on the number of iterations and other system
parameters. In our current implementation we assume that user can specify 
which query submission statements to be transformed. Making this decision
in a cost-based manner is a topic of future work.

\eat{
\subsection*{Limiting the number of iterations before fetch}
The program transformations presented in this paper, specifically
Rule~\ref{fig:rule-fission}, is designed such that the first call to
{\em fetchResult} is made only after all the loop iterations, which make the
asynchronous query submission, are completed. This may not be a good idea in
general. It is possible to extend our transformation rules so as to allow the
second loop (which consumes the query results) to begin after a specific
number of asynchronous query submissions. This can be achieved by 
enclosing the two loops generated after the fission into a parent loop.
We omit the details of this extension from this paper.
}


%


\section{Conclusion} \label{sec:concl}
We propose a program analysis and transformation based approach to
automatically rewrite database applications to exploit the benefits of
asynchronous query submission. The techniques
presented in this paper significantly increase the applicability of known
techniques to address this problem.  We also described a novel approach
to combine asynchronous submission with our earlier work on batching in
order to achieve a balance between the trade offs of batching and
asynchronous query submission.

Although our program transformations are
presented in the context of database queries, the techniques are general in
their applicability, and can be used in other contexts such as calls to Web
services, as shown by our experiments.  We presented a detailed experimental study, carried out on
real-world and publicly available benchmark applications. Our experimental
results show performance gains to the extent of 75\% in several cases. 
Finally, we identify some interesting directions along which this work can be extended.

\vspace{3mm}

\noindent\textbf{Acknowledgements:}
We thank Yatish Turakhia for his help in the implementation of asynchronous
batching and integrating the ehcache library into DBridge.

\ifCLASSOPTIONcaptionsoff
  \newpage
\fi



\bibliographystyle{IEEEtran}
\bibliography{jasync.bib}

\begin{thebibliography}{10}
\providecommand{\url}[1]{#1}
\csname url@rmstyle\endcsname
\providecommand{\newblock}{\relax}
\providecommand{\bibinfo}[2]{#2}
\providecommand\BIBentrySTDinterwordspacing{\spaceskip=0pt\relax}
\providecommand\BIBentryALTinterwordstretchfactor{4}
\providecommand\BIBentryALTinterwordspacing{\spaceskip=\fontdimen2\font plus
\BIBentryALTinterwordstretchfactor\fontdimen3\font minus
  \fontdimen4\font\relax}
\providecommand\BIBforeignlanguage[2]{{%
\expandafter\ifx\csname l@#1\endcsname\relax
\typeout{** WARNING: IEEEtran.bst: No hyphenation pattern has been}%
\typeout{** loaded for the language `#1'. Using the pattern for}%
\typeout{** the default language instead.}%
\else
\language=\csname l@#1\endcsname
\fi
#2}}

\bibitem{GUR08}
R.~Guravannavar and S.~Sudarshan, ``{R}ewriting {P}rocedures {f}or {B}atched
  {B}indings,'' in \emph{Intl. Conf. on Very Large Databases}, 2008.

\bibitem{DBR}
M.~Chavan, R.~Guravannavar, K.~Ramachandra, and S.~Sudarshan, ``Dbridge: A
  program rewrite tool for set-oriented query execution,'' in \emph{ICDE},
  2011, pp. 1284--1287.

\bibitem{SOAP12}
K.~Ramachandra, R.~Guravannavar, and S.~Sudarshan, ``Program analysis and
  transformation for holistic optimization of database applications,'' in
  \emph{ACM SIGPLAN SOAP 2012}, 2012, pp. 39--44.

\bibitem{CHA11}
M.~Chavan, R.~Guravannavar, K.~Ramachandra, and S.~Sudarshan, ``Program
  transformations for asynchronous query submission,'' in \emph{ICDE}, 2011,
  pp. 375--386.

\bibitem{GUR09}
R.~Guravannavar, ``Optimization and evaluation of nested queries and
  procedures,'' {Ph.D.} Thesis, IIT Bombay, 2009.

\bibitem{MUCHNICK}
S.~S. Muchnick, \emph{{Advanced Compiler Design and Implementation}}.\hskip 1em
  plus 0.5em minus 0.4em\relax San Francisco, CA, USA: Morgan Kaufmann
  Publishers Inc., 1997.

\bibitem{PDG87}
J.~Ferrante, K.~J. Ottenstein, and J.~D. Warren, ``{T}he {P}rogram {D}ependence
  {G}raph and {I}ts {U}se in {O}ptimization,'' \emph{ACM Trans. Program. Lang.
  Syst.}, vol.~9, no.~3, pp. 319--349, 1987.

\bibitem{KENEDY90}
K.~Kennedy and K.~S. McKinley, ``{L}oop {D}istribution with {A}rbitrary
  {C}ontrol {F}low,'' in \emph{Proceedings of Supercomputing}, 1990.

\bibitem{SOOT}
``{Soot: A Java Optimization Framework \\ http://www.sable.mcgill.ca/soot}.''

\bibitem{EXEC}
\BIBentryALTinterwordspacing
``The java executors framework.'' [Online]. Available:
  \url{http://docs.oracle.com/javase/tutorial/essential/concurrency/executors.%
html}
\BIBentrySTDinterwordspacing

\bibitem{AMIT09}
A.~Manjhi, C.~Garrod, B.~M. Maggs, T.~C. Mowry, and A.~Tomasic, ``{H}olistic
  {Q}uery {T}ransformations for {D}ynamic {W}eb {A}pplications,'' in
  \emph{Intl. Conf. on Data Engineering}, 2009.

\bibitem{EHCACHE}
\BIBentryALTinterwordspacing
``The ehcache java caching library.'' [Online]. Available:
  \url{http://ehcache.org/}
\BIBentrySTDinterwordspacing

\bibitem{JMOB}
\BIBentryALTinterwordspacing
``{ObjectWeb Consortium-JMOB (Java middleware open benchmarking)}.'' [Online].
  Available: \url{http://jmob.ow2.org/}
\BIBentrySTDinterwordspacing

\bibitem{FREEBASE}
``{The Freebase repository: http://www.freebase.com/}.''

\bibitem{LINUXAIO}
``{Kernel Asynchronous I/O (AIO) Support for Linux \\
  http://lse.sourceforge.net/io/aio.html}.''

\bibitem{GRA03}
G.~Graefe, ``{E}xecuting {N}ested {Q}ueries,'' in \emph{10th Conference on
  Database Systems for Business, Technology and the Web}, 2003.

\bibitem{GAL07}
M.~Elhemali, C.~A. Galindo-Legaria, T.~Grabs, and M.~M. Joshi, ``{E}xecution
  {S}trategies for {SQL} {S}ubqueries,'' in \emph{ACM SIGMOD}, 2007.

\bibitem{IYENGAR08}
S.~Iyengar, S.~Sudarshan, S.~Kumar, and R.~Agrawal, ``{E}xploiting
  {A}synchronous {IO} using the {A}synchronous {I}terator {M}odel,'' in
  \emph{Intl. Conf. on Management of Data (COMAD)}, 2008.

\bibitem{STEVEN00}
S.~P. Vanderwiel and D.~J. Lilja, ``{Data Prefetch Mechanisms},'' \emph{ACM
  Computing Surveys}, vol.~32, no.~2, 2000.

\bibitem{KWOK04}
K.~C. Yeung, ``{Dynamic Performance Optimisation of Distributed Java
  Applications},'' Ph.D. dissertation, Imperial College of Science, Technology
  and Medicine, 2004.

\bibitem{Arjun09}
A.~Dasgupta, V.~Narasayya, and M.~Syamala, ``A static analysis framework for
  database applications,'' in \emph{ICDE '09}, 2009, pp. 1403--1414.

\bibitem{Surajit09}
S.~Chaudhuri, V.~Narasayya, and M.~Syamala, ``Bridging the application and dbms
  divide using static analysis and dynamic profiling,'' in \emph{SIGMOD}, 2009,
  pp. 1039--1042.

\bibitem{BowmanS07}
I.~T. Bowman and K.~Salem, ``Semantic prefetching of correlated query
  sequences,'' in \emph{ICDE}, 2007.

\bibitem{statusquo}
A.~Cheung, O.~Arden, S.~M.~A. Solar-Lezama, and A.~C. Myers, ``Statusquo:
  Making familiar abstractions perform using program analysis,'' in
  \emph{CIDR}, 2013.

\bibitem{AMIT08}
A.~Manjhi, ``{Increasing the Scalability of Dynamic Web Applications},'' Ph.D.
  dissertation, Carnegie Mellon University, 2008.

\bibitem{RAM12}
K.~Ramachandra and S.~Sudarshan, ``Holistic optimization by prefetching query
  results,'' in \emph{ACM SIGMOD 2012}, 2012, pp. 133--144.

\bibitem{HIB}
\BIBentryALTinterwordspacing
``{Hibernate OR Mapping tool}.'' [Online]. Available:
  \url{http://www.hibernate.org}
\BIBentrySTDinterwordspacing

\bibitem{Par96}
S.~Parthasarathy, W.~Li, M.~Cierniak, and M.~J. Zaki, ``Compile-time
  inter-query dependence analysis,'' in \emph{IEEE Symp. on Parallel and
  Distr.\ Processing}, 1996.

\end{thebibliography}
\end{document}